\documentclass[12pt,a4paper]{article}

\usepackage{graphicx}
\usepackage{subfigure}

\usepackage{amsmath}
\usepackage{amssymb}

\begin{document}

\newcommand{\mb}[1]{\mathbf{#1}}

\title{\bf{Energy localization in two chaotically coupled systems}}
\author{Johan Gr\"onqvist\footnote{e-mail: Johan.Gronqvist@matfys.lth.se} \
 and Thomas Guhr\footnote{e-mail: Thomas.Guhr@matfys.lth.se} \\ \\
{\it Mathematical Physics, LTH, Lund University, Lund, Sweden}}
\date{\today}

\maketitle

\begin{abstract}
We set up and analyze a random matrix model to study energy
localization and its time behavior in two chaotically coupled
systems. This investigation is prompted by a recent experimental and
theoretical study of Weaver and Lobkis on coupled elastomechanical
systems. Our random matrix model properly describes the main features
of the findings by Weaver and Lobkis. Due to its general character,
our model is also applicable to similar systems in other areas of
physics -- for example, to chaotically coupled quantum dots.
\end{abstract}

\section{Introduction}

The statistical features of coupled systems have attracted considerable
interest in many branches of physics. Random matrix theory (RMT) has
been successfully used in many of those investigations. RMT was
founded by Wigner~\cite{Wigner}. It is a schematic model~\cite{Mehta}
in which the Hamiltonian, or, more generally, the wave operator of the
system, is replaced by a random matrix. The necessary prerequisite is
that the system be sufficiently ``complex,'' implying that the matrix
elements of the Hamiltonian, or wave operator, calculated in an
arbitrary basis, behave like random numbers. It has been shown that
the spectral fluctuations in numerous different systems, if measured
on the scale of the local mean-level spacing, are very well modeled by
RMT; see the reviews in Refs.~\cite{GMGW,Stockmann,Haake}. Due to the
connection with chaos, one frequently refers to those systems as
quantum chaotic which show correlations of RMT type.  Similarly,
systems are often referred to as regular if they lack spectral
correlations.

We consider two coupled systems. We assume that either the two systems
are chaotic before they are coupled or that the coupling itself
introduces chaoticity if the separate systems are regular. This
scenario is equivalent to the breaking of symmetries, if only two
values of the quantum number belonging to that symmetry are taken into
account.  The statistical features crucially depend on the strength of
the coupling measured on the scale of the local mean-level spacing.
Many studies have been devoted to this issue of chaotically coupled
systems or, equivalently, to symmetry breaking. We mention isospin
breaking in nuclear physics~\cite{values,GW}, symmetry breaking in
molecular physics~\cite{Leitner}, symmetry breaking in resonating
quartz crystals~\cite{cristais2} and coupled microwave
billiards~\cite{nos}. While these studies addressed the spectral
correlation, several investigations in nuclear
physics~\cite{26Al,22Na,BGH,HNSA} focused on the statistics of the
wave functions and related observables in the presence of symmetry
breaking or similar effects.  In all these cases, RMT approaches in
the spirit of the Rosenzweig-Porter model~\cite{Porter} were
successful.

Sometimes observables in the time domain such as spectral form factors
are more appropriate than the eigenenergy correlations
functions~\cite{Mehta,GMGW,Stockmann,Haake}. This is so, for example,
in the case of the presently much discussed fidelity; see
Refs.~\cite{CT,BCV,PSZ,VH} and references therein. Another example is
the study of the energy spread in chaotic
systems~\cite{cohen00,kottos01}.  In the context of coupled systems,
the time evolution of wave packets was investigated in
Ref.~\cite{BTU}.

In the present contribution, we study energy localization in two
coupled systems in the time domain. This problem was addressed in a
recent work by Weaver and Lobkis~\cite{WL} who measured the time
dependence of the wave intensity distribution in two coupled
reverberation rooms. To this end, these authors recorded the time
response to an elastic excitation of two coupled aluminum cubes.
Moreover, they investigated the same problem theoretically and they
numerically calculated the response in coupled two-dimensional
membranes. In our study, we se tup and analyze an RMT model, based on
the approaches in Refs.~\cite{Porter,GW,BGH}. Its general character
makes our model useful for similar problems in different physics
contexts. In particular, we expect that our RMT approach also applies
to coupled quantum dots.

The article is organized as follows. In Sec.~\ref{sec:exp_and_model}
we sketch the work of Weaver and Lobkis~\cite{WL}. In
Sec.~\ref{sec:RMT} we set up the RMT model and analyze it analytically
and numerically.  We compare our results to those of Weaver and Lobkis
in Sec.~\ref{sec:RMT_result}. Discussion and conclusions are given in
Sec.~\ref{sec:dic_concl}.

\section{Experiment and numerical calculations}
\label{sec:exp_and_model}

\begin{figure}
  \centering 
  \subfigure[Low frequency
  bins.]{\includegraphics[height=4cm]{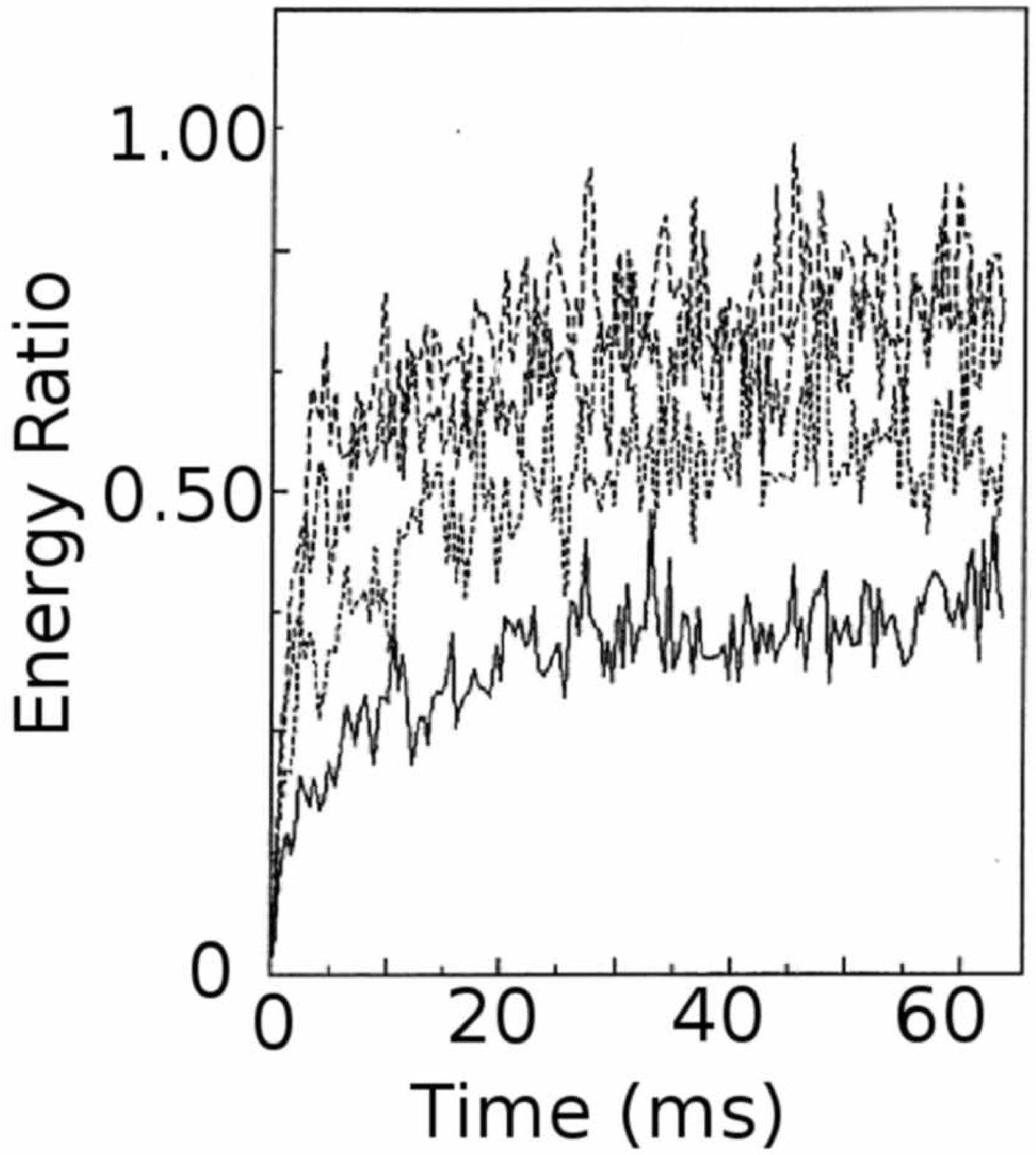} 
  \label{fig:exp_loc}} \quad 
  \subfigure[High frequency bins.]{\includegraphics[height=4cm]{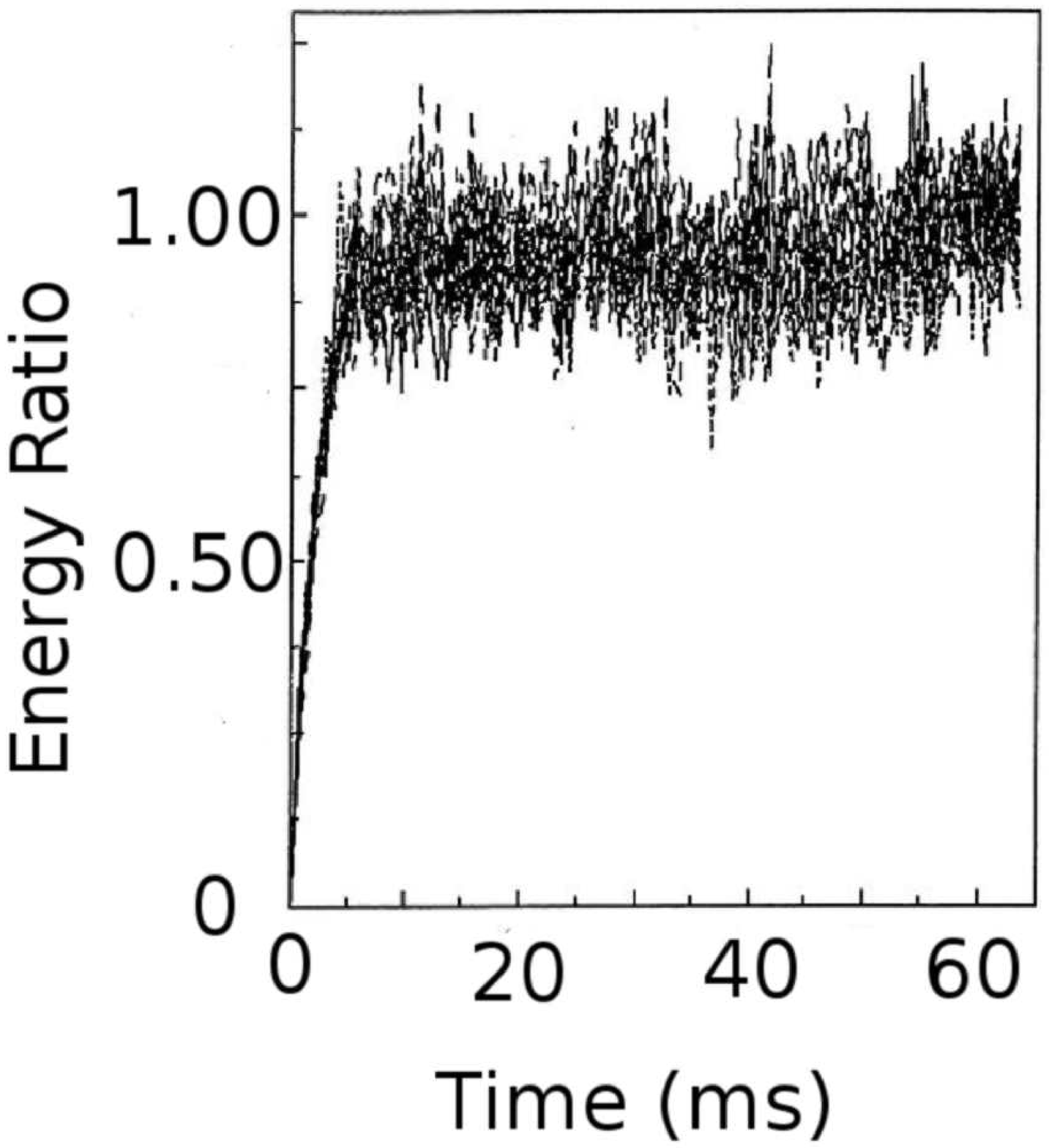}\label{fig:exp_nonlocalised}} 
  \caption{Results from the experiment by Weaver and Lobkis. The
  energy ratio between the different rooms is plotted versus time (in
  ms). We note the different
  scales on the $y$ axes. Reprinted from Ref.~\cite{WL}.}   
\label{fig:WL_exp}
\end{figure}

As we aim at a comparison with their findings, we present the work of
Weaver and Lobkis~\cite{WL} in some detail. Thereby, we also introduce
the notation and conventions. The system studied experimentally consists
of two aluminum cubes coupled by a solid connection, manufactured out
of a solid aluminum block.  The corners of the cubes were removed to
desymmetrize the structure. This was done to ensure ``chaotic''
motion. Elastomechanical wave modes were excited in one room, and the
response was measured in the other room. In this way, 16
different curves of energy intensity versus time were recorded, each
in a small region around a different frequency. The results show that
the energy does not always spread equally over the two rooms. If the
coupling is weak, then the wave intensity is higher in the room where
the initial excitation was performed than in the other room,
regardless of how long one waits. Hence, the energy ratio never
approaches unity. This deviation from the equipartition of the energy
in the two rooms is referred to as energy localization. The resulting
data are shown in Fig.~\ref{fig:WL_exp}, and as expected there is
localization in the bins of larger mean-level spacing, but not in the
bins of small mean-level spacing. We will discuss these results
further in Sec.~\ref{sec:RMT_result}.

Two numerical studies were performed on membranes with rough
boundaries~\cite{WL}. The dynamics of the system was in both cases
governed by discretized wave equations in each of the rooms, and the
coupling was realized in different ways in the two numerical
calculations, to which we refer as N1 and N2 in the sequel. In the
first one, N1, the connection had the form of a window between the two
membranes similar to the situation in the experimental setup. In the
second numerical calculation, N2, the rooms were separated, but
springs were attached to a few different sites in rooms 1 and 
2, thereby coupling those sites. In N1 and N2, a nonvanishing
initial condition was given to one site in one of the rooms, and the
response was calculated at different sites in the other room. The
resulting time series were cosine-bell time windowed to focus on a
specific instant in time. Then, the time series were Fourier
transformed and integrated over a small region in frequency to
accumulate data around a certain frequency. As in the experiment,
16 different curves of intensity versus time were obtained around
different frequencies. At different frequencies, the systems have
different effective couplings. Therefore, one expects~\cite{WL} the
time behavior of the different curves to differ in the degree of
localization, as well as in the way in which this asymptotic
saturation value is reached. Due to the differences in the coupling
mechanism, there will also be differences between the results of N1
and N2.

Moreover, Weaver and Lobkis performed an analytical model study.  The
elastic wave equation for the state of the system $U(t)$, say, is of
second order in time. To focus on the response in a narrow interval
around a certain frequency $\Omega$, the ansatz $U(t) =
u(t)\exp(-i\Omega t)$ is made with the assumption that $u(t)$ varies
slowly with time. This leads to a first-order differential equation in
time for $u(t)$ which has the form
\begin{equation}
-i\frac{\partial}{\partial t} u(t) = \frac{C+\Omega^2}{2 \Omega} u(t) \ , 
\label{eq:H_v_def}
\end{equation}   
where $C$ is the wave operator of the original second-order equation.
The energies $E_1(t)$ and $E_2(t)$ in rooms 1 and 2 at time $t$
are defined as the total probability density of finding the system
state $u(t)$ in one of the states $\psi_{ik}, \ i=1,2$, which are good
eigenstates in room $i$:
\begin{equation}
E_i(t) = \sum_k \left| u(t) \cdot \psi_{ik} \right| ^2 \ .
\end{equation}
Strictly speaking, $E_i(t)$ is no energy. Nevertheless, we find this
terminology introduced in Ref.~\cite{WL} appropriate and use it as
well, because $E_i(t)$ measures the degree of motion in room $i$.  If
no energy dissipates into the surrounding environment, the total
energy $E=E_1(t)+E_2(t)$ is conserved -- i.e., independent of time. In
the sequel, it is always assumed that the system is excited in room
1, and the energy is measured in room 2.  Analytical solutions for
two coupled states are presented in Ref.~\cite{WL} by employing different
statistical assumptions.

\section{Random matrix model}
\label{sec:RMT}

We set up the model in Sec.~\ref{sec:RMT_model}. The connection to the
two-level form factor is established in Sec.~\ref{RMTff}, and a
$2\times 2$ version of the model is evaluated in Sec.~\ref{M22}.  We
discuss numerical simulations of the RMT model in
Sec.~\ref{sec:RMT_num}. Finally, we comment on chaotically coupled
regular systems in Sec.~\ref{sec:RMT_pos}.

\subsection{Setup of the model}
\label{sec:RMT_model}

Spectral correlations in elastomechanics have been shown to be well
described by RMT \cite{Weaver}. This is also true in the case of
symmetry breaking~\cite{cristais2}, which is of direct relevance for
the present study. Thus, RMT is also likely to be capable of modeling
the time behavior of elastomechanical systems.  As the first-order
equation~\eqref{eq:H_v_def} has proved to be a good approximation to
the experimental situation, we also base our model on this
Schr\"odinger type of equation. Thus, it is more natural to
replace $(C+\Omega^2)/2\Omega$ by the random matrix $H$ than to
replace $C$ itself. It turns out that this is indeed the best choice.

The appropriate RMT model is an extension of the one employed in
Refs.~\cite{GW,BGH}. The random matrix $H$ modeling the operator
$(C+\Omega^2)/2\Omega$ reads
\begin{equation}
H = \left[ \begin{array}{cc} H_1 & 0 \\ 
                               0 & H_2 \end{array} \right] + 
    \alpha \left[ \begin{array}{cc}  0 & V \\ 
                                     V^\dagger & 0 \end{array} \right] \ .
\label{eq:RMT_model_def}
\end{equation}  
The two matrices $H_i, \ i=1,2$, model the uncoupled rooms 1 and
2. They are real symmetric and have random entries. We draw them
from (two independent) Gaussian orthogonal ensembles (GOE's). As the
rooms in the experiments and the numerical calculations N1 and N2 were
of the same size, the level densities were also the same. This can be
adjusted in the RMT model by giving the matrices $H_i$ the same
statistical weights and the same dimension $N$, such that $2N$ is the
total dimension of $H$. The matrix $V$ also has random entries. The
strength of the coupling is measured by the dimensionless parameter
$\alpha$. It is sufficient to always assume $\alpha \ge 0$. The
statistical weight of $V$ is chosen such that the total $H$ is in the
GOE of $2N\times 2N$ matrices for $\alpha = 1$.

We write the eigenvalue equation for the total Hamiltonian $H$ in the form
\begin{eqnarray}
H\Psi_{n} = \omega_{n}\Psi_{n} \ , \qquad n=1,\ldots,2N \ .
\label{evh}
\end{eqnarray}
The eigenvalues $\omega_{n}$ and eigenvectors $\Psi_{n}$ are functions
of the coupling parameter $\alpha$. It is convenient to introduce the
notation
\begin{eqnarray}
\Psi_{n} =  \left[ \begin{array}{c} \Psi_{1n} \\ 
                                    \Psi_{2n} \end{array} \right] \ ,
\label{psi12}
\end{eqnarray}
where $\Psi_{in}, \ i=1,2$, is the projection of $\Psi_n$ onto the
subspace $i$. We emphasize that $\Psi_{in}, \ i=1,2$, are
functions of the coupling parameter $\alpha$. The eigenvalue equations
for the Hamiltonians $H_i$ are written as
\begin{eqnarray}
H_1\psi_{1n} &=& \omega_{1n}\psi_{1n} \ , \qquad n=1,\ldots,N \ , \nonumber\\
H_2\psi_{2n} &=& \omega_{2n}\psi_{2n} \ , \qquad n=1,\ldots,N \ ,
\label{evhi}
\end{eqnarray}
where the eigenvalues $\omega_{in}$ and eigenvectors $\psi_{in}$ are
not functions of $\alpha$. This difference between $\Psi_{in}$ and
 $\psi_{in}$ is an immediate consequence of the fact that $H$
depends on $\alpha$, while $H_1$ and $H_2$ do not. We will also use
the notation $\widehat{\psi}_{2n}=(0,\psi_{2n})^\dagger$ for the
corresponding $2N$-dimensional vector with zeros in the first $N$
components.

At time $t=0$, the system is excited in room one such that the state
of the total system can be written as
\begin{eqnarray}
S =  \left[ \begin{array}{c} s \\ 
                             0 \end{array} \right] \ ,
\label{source}
\end{eqnarray}
where $s$ in room one is not specified in detail. We refer to $S$ as
to the source. The time evolution of the source is then simply
\begin{eqnarray}
u(t) = T(t)S \ , \qquad {\rm where} \qquad
T(t) = \exp\left(iHt\right)
\label{teo}
\end{eqnarray}
is the time evolution operator.  Using the eigenvectors $\widehat{\psi}_{2n}$,
the energy in room 2 is thus given by
\begin{eqnarray}
E_2(t,\alpha) &=& \sum_{n=1}^{N} \left| u(t) \cdot \widehat{\psi}_{2n} \right|^2 
        = \sum_{n=1}^{N} \left( T(t)S \right) ^{\dagger}\widehat{\psi}_{2n}
             \widehat{\psi}_{2n}^\dagger \left( T(t)S \right) \nonumber \\
       &=& S^{\dagger} T^{\dagger}(t) 
             \left[\begin{array}{cc} 0 & 0\\
                                     0 & 1_N \end{array} \right]
             T(t) S \ .
\label{eq:RMT_E2}
\end{eqnarray}
The block matrix only contains the unit matrix $1_N$ for room 2.  We
write $E_2(t,\alpha)$ instead of $E_2(t)$ to emphasize the dependence
on the coupling parameter $\alpha$ in the RMT model. We will study
averages $\overline{E_2}(t,\alpha)$ over the ensemble of matrices
introduced in Eq.~\eqref{eq:RMT_model_def}.  Occasionally, we will
also consider averages over the direction of the source. This average
is denoted by angular brackets $\langle \cdots \rangle$.

The total energy $E$ is trivially conserved in the framework of our
model. We have
\begin{eqnarray}
E = E_1(t,\alpha)+E_2(t,\alpha) = S \cdot S = s \cdot s \ .
\label{etot}
\end{eqnarray}
Hence, we can always construct $E_1(t,\alpha)$, once $E_2(t,\alpha)$
has been calculated.

As all correlations have to be
measured~\cite{Mehta,GMGW,Stockmann,Haake} on the local scale of the
mean-level spacing $D$, we introduce an unfolded time $\tau=Dt$. This
is also the scale on which $\alpha$ acts; we introduce the unfolded
coupling parameter $\lambda=\alpha/D$. The energy on the unfolded
scale is then given by
\begin{eqnarray}
\varepsilon_2(\tau,\lambda) =
            \lim_{N\to\infty} \overline{E_2}(\tau/D,D\lambda).
\label{eun}
\end{eqnarray}
Small values of $\alpha$ have a large effect on the correlation
functions if the mean-level spacing $D$ is small as well.

\subsection{Relation to the two-level form factor}
\label{RMTff}

We now derive an estimate for the time evolution of $E_2(t,\alpha)$
which should apply to strong coupling strength-- i.e., to a parameter
$\alpha$ which is large on the unfolded scale of the mean level
spacing. If one assumes that the source $S$ comprises excitations into
all states $\psi_{1n}$, one may average over the direction of the
source. We find, from Eq.~\eqref{eq:RMT_E2},
\begin{eqnarray}
\langle E_2\rangle (t,\alpha) = B {\rm tr}^{(11)}\,T^{\dagger}(t) 
             \left[\begin{array}{cc} 0 & 0\\
                                     0 & 1_N \end{array} \right] T(t) \ ,
\label{e2est1}
\end{eqnarray}
where ${\rm tr}^{(11)}$ is the trace over the $(11)$ sector of the
whole matrix -- i.e., over the upper left block. The constant $B$ results
from the average. Expanding the time evolution operator in terms of
the eigenvectors,
\begin{eqnarray}
T(t) = \sum_{n=1}^{2N} \Psi_n \exp\left(i\omega_nt\right) \Psi_n^\dagger \ ,
\label{teoexp}
\end{eqnarray}
we arrive after a short calculation at
\begin{eqnarray}
\langle E_2\rangle (t,\alpha) = B \sum_{n,m} 
                        \left(\Psi_{1m}\cdot\Psi_{1n}\right)
                        \left(\Psi_{2n}\cdot\Psi_{2m}\right)
                        \exp\left[i(\omega_n-\omega_m)t\right] \ .
\label{e2est2}
\end{eqnarray}
The vectors $\Psi_{1n}$ and $\Psi_{2n}$ are, according to
Eq.~\eqref{psi12}, the projections of the full eigenvector $\Psi_{n}$
onto the subspaces corresponding to the two rooms. They depend on the
coupling parameter $\alpha$. These vectors coincide with the
eigenvectors for the matrices $H_1$ and $H_2$ only for $\alpha=0$.
Hence, they are only orthogonal in this special case. For all values
$\alpha>0$, the scalar products $\Psi_{1m}\cdot\Psi_{1n}$ and
$\Psi_{2n}\cdot\Psi_{2m}$ are neither zero nor given by $\delta_{nm}$.
The ensemble average over the eigenenergies $\omega_n$ and over the
eigenfunctions $\Psi_{n}$ decouples only for the parameter values
$\alpha=0$ and $\alpha=1$-- i.e., if the system either falls into two
GOE's or is represented by one GOE. We now estimate the ensemble average
of the energy $\langle E_2\rangle (t,\alpha)$ by making the
approximation that the averages over eigenenergies and eigenfunctions
decouple. For strong coupling this should yield a reasonable result,
and we will use this approximation in that limit only.  We average
over the eigenfunctions. For each term in the double sum in
Eq.~\eqref{e2est2} the product of scalar products gives the same
contribution which we can take out of the sums. Thus, we find
\begin{eqnarray}
\overline{\langle E_2\rangle} (t,\alpha) 
 = B \overline{\sum_{n,m}\exp\left[i(\omega_n-\omega_m)t\right]} \ ,
\label{e2est3}
\end{eqnarray}
where we absorbed the contributions from the average over the scalar
products into the constant $B$. The average over the
eigenvalues is yet to be performed.  Luckily, the average in
Eq.~\eqref{e2est3} is recognized as the Fourier transform of the
two--level correlation function~\cite{Mehta,GMGW}. This yields, on the
unfolded scale,
\begin{eqnarray}
\varepsilon_2(\tau,\lambda) = B \left[\delta(\tau/2\pi) 
                            + K_2(\tau/2\pi,\lambda)\right] \ ,
\label{e2est4}
\end{eqnarray}
where the function $K_2(\tau,\lambda)$ is referred to as the two-level
form factor. The term $\delta(\tau)$ is due to the diagonal
contributions in the double sum of Eq.~\eqref{e2est3}. We notice that
the conventions used here require a rescaling of time with a factor of
$2\pi$.  The expression~\eqref{e2est4} is an estimate for the energy
$\varepsilon_2(\tau,\lambda)$, exclusively in terms of the 
Fourier-transformed two-level spectral correlation function for the
transition from two GOE's to one GOE. Unfortunately, the two-level form
factor is not known analytically for all values of $\lambda$.  The
result for one GOE, corresponding to $\lambda\to\infty$,
reads~\cite{Mehta}
\begin{equation}
K_2(\tau,\infty) = \left\{ 
            \begin{array}{ll} 2\tau-\tau \ln(2\tau+1) & 
                                   {\rm for} \quad 0<\tau\le 1 , \\ 
                              2-\tau \ln\frac{{\textstyle 2\tau+1}}
                                             {{\textstyle 2\tau-1}} &
                                   {\rm for} \quad \tau\ge 1 .
            \end{array} \right. 
\label{eq:formfactor_result}
\end{equation} 
We notice that the function $\varepsilon_2(t,\infty)$ has the
limit properties
\begin{eqnarray}
\varepsilon_2(0,\infty) &=& 0 , \nonumber \\
\lim _{\tau \rightarrow \infty} \varepsilon_2(\tau,\infty) &=& B \ .
\label{eq:ff_props}
\end{eqnarray}
The second property will be used to fix the scale in comparison with
the results of Ref.~\cite{WL}.  The estimate~\eqref{e2est4} should
apply to large coupling -- i.e., to parameter values $\lambda\gg 1$.

\subsection{Two-by-two model}
\label{M22}

Quite often, one obtains surprisingly good information about an RMT
model by restricting it to the smallest possible matrix dimension such
that the nontrivial specific characteristics of the model are still
present. This is successful in the case of the nearest-neighbor
spacing distribution for the Gaussian orthogonal, unitary and
symplectic ensembles (GGOE, GUE and GSE, respectively) ~\cite{GMGW,Stockmann,Haake}, but also for crossover
transitions~\cite{lenz}. Here, we proceed analogously.  We obtain a
$2\times 2$ RMT model by setting $N=1$ in
Eq.~\eqref{eq:RMT_model_def}.  It turns out convenient to absorb the
parameter $\alpha$ into the matrix element $V$ such that
\begin{equation}
  H = \left[ \begin{array}{cc} H_1 & V \\ V & H_2 \end{array} \right]
\label{m1}
\end{equation} 
and to readjust the probability density function $P(H)$ for the
ensemble average accordingly:
\begin{equation}
  P(H) = \frac{\sqrt{2}}{\pi\sqrt{\pi\alpha^2}} 
               \exp\left(-H_1^2-H_2^2-2\frac{V^2}{\alpha^2}\right) \ .
\label{m2}
\end{equation} 
As before, the GOE is recovered for $\alpha=1$. We introduce
eigenvalue and angle coordinates
\begin{equation}
  \left[ \begin{array}{cc} H_1 & V \\ 
                             V & H_2 \end{array} \right] = 
  \left[ \begin{array}{cc} \cos\varphi & -\sin\varphi \\ 
                           \sin\varphi &  \cos\varphi \end{array} \right]
  \left[ \begin{array}{cc} \omega_1 & 0 \\ 
                                  0 & \omega_2 \end{array} \right] 
  \left[ \begin{array}{cc} \cos\varphi & \sin\varphi \\ 
                          -\sin\varphi & \cos\varphi \end{array} \right] \ ,
\label{m3}
\end{equation} 
which implies that the time evolution can be written in the same form
with eigenvalues $\exp(i\omega_1t)$ and $\exp(i\omega_2t)$. For the
integration measure one has
\begin{equation}
dH_1 dH_2 dV = \frac{1}{4} |\omega_1-\omega_2| d\omega_1 d\omega_2 d\varphi \ .
\label{m4}
\end{equation} 
The source in this $2\times 2$ model is simply given by $S=(1,0)$.
Thus, collecting everything, we find, with Eq.~\eqref{eq:RMT_E2},
\begin{eqnarray}
\overline{E_2}(t,\alpha) &=& \frac{\sqrt{2}}{4\pi\sqrt{\pi\alpha^2}}
             \int\limits_{-\infty}^{+\infty}d\omega_1
             \int\limits_{-\infty}^{+\infty}d\omega_2|\omega_1-\omega_2|   
             \int\limits_0^{2 \pi} d\varphi \nonumber\\
    & & \qquad \exp\left[-\frac{(\omega_1+\omega_2)^2}{2}-
        \frac{(\omega_1-\omega_2)^2}{2} \left(\cos ^2 2\varphi + \frac{\sin ^2
             2\varphi}{\alpha^2} \right) \right] \nonumber \\ 
    & & \qquad \sin^22\varphi \, \sin^2\frac{\omega_1-\omega_2}{2}t \ .
\label{m5}
\end{eqnarray}
We introduce $x=(\omega_1+\omega_2)/\sqrt{2}$ and
$y=(\omega_1-\omega_2)/\sqrt{2}$ as new integration variables, perform
the $x$ and $\varphi$ integrals, and arrive at
\begin{eqnarray}
\overline{E_2}(t,\alpha) &=& \frac{\alpha}{2 (\alpha + 1)} - 
 \nonumber\\
 & & \qquad \alpha \int\limits_0^\infty dy\, y \, 
                   \exp\left[-y^2(1+\alpha^2)\right] 
 \nonumber\\
 & & \qquad \qquad 
\left[{\rm I}_0 (y^2 (1-\alpha^2)) - {\rm I}_1(y^2 (1-\alpha^2))\right] 
 \cos 2\alpha yt \, .
\label{eq:2d_result}
\end{eqnarray}
Here, ${\rm I}_0$ and $\rm{I}_1$ denote the modified Bessel functions
of zeroth and first order, respectively.  

Formula~\eqref{eq:2d_result} is the final result in the framework of
our $2\times 2$ model. In Ref.~\cite{WL}, solutions for two-state
models based on Eq.~\eqref{eq:H_v_def} with different statistical
assumptions were given. Although the general behavior is similar to
our $2\times 2$ RMT result~\eqref{eq:2d_result}, the analytical forms
of these solutions in Ref.~\cite{WL} are different.

For obvious reasons, the unfolding of formula~\eqref{eq:2d_result} is
meaningless.  Nevertheless, experience with $2\times 2$ model for the
spacing distribution tells that meaningful statements can be achieved
if the transition parameter -- i.e., the coupling strength $\alpha$ in
our case -- is interpreted properly.  Here, we can do that in the
following manner. For large times $t$, the function
$\overline{E_2}(t,\alpha)$ becomes constant, because it reaches its
saturation limit. The latter will depend on $\alpha$. Thus, comparing
the saturation limit for the $2\times 2$ model with that of the
$2N\times 2N$ RMT simulation or with those of the experiment and the
numerical calculations N1 and N2 allows one, in principle, to
interpret $\alpha$ on the unfolded scale.  The saturation limit is
easily obtained from Eq.~\eqref{eq:2d_result}.  Due to the
Riemann-Lebesgue lemma~\cite{GR}, we have
\begin{eqnarray}
\lim_{t\to\infty}\overline{E_2}(t,\alpha) = \frac{\alpha}{2 (\alpha + 1)} \ ,
\label{m7} 
\end{eqnarray}
because the cosine function in the integrand oscillates so rapidly
for large $t$, that the integral gives zero in the limit $t
\rightarrow \infty$.

We notice that there is no equipartition of the energies for
$t\to\infty$ at $\alpha = 1$. This may seem a bit unexpected, because
$H$ is a $2\times 2$ GOE matrix for that parameter value. It simply
reflects the need to properly interpret the parameters, as just
discussed. We study a two-state system and compare the probability
for a transition between the two states with the probability of
staying in one state. There is no reason why these should be
equal. In the $N \rightarrow \infty$ limit, however, we expect
equipartition for the full GOE. As the coupling parameter has to be
measured on the scale of the local mean level spacing, the limit $N
\rightarrow \infty$ corresponds to the limit $\alpha \rightarrow \infty$ in
the $2\times 2$ version, and in that limit the right-hand side of
Eq.~\eqref{m7} tends to $1/2$ and therefore to equipartition.

\begin{figure}[htb]
  \centering
  \includegraphics[height=4cm]{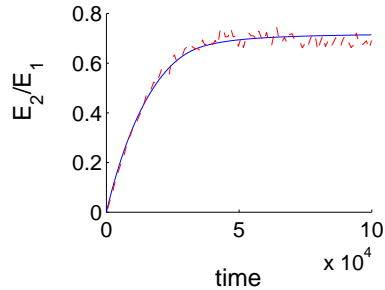}
  \caption{Energy ratio $E_2 / E_1$ versus (dimensionless) time. Data
  from N1 of Ref.~\cite{WL}, bin 16, plotted
  as dashed line. The result from 
    Eq.~\eqref{e2est4} for $\lambda \rightarrow \infty$, $B$ set by property 2
    of Eq.~\eqref{eq:ff_props} and a rescaling of time to make the two curves
    fit for early times, is plotted as a solid line.}
  \label{fig:WL_num1_vs_form_factor}
\end{figure} 

\begin{figure}[htb]
  \centering \subfigure[$2 \times 2$ RMT
  model, fitted to N2 bins 2,3,4 from top to bottom.]{\includegraphics[height=4cm]{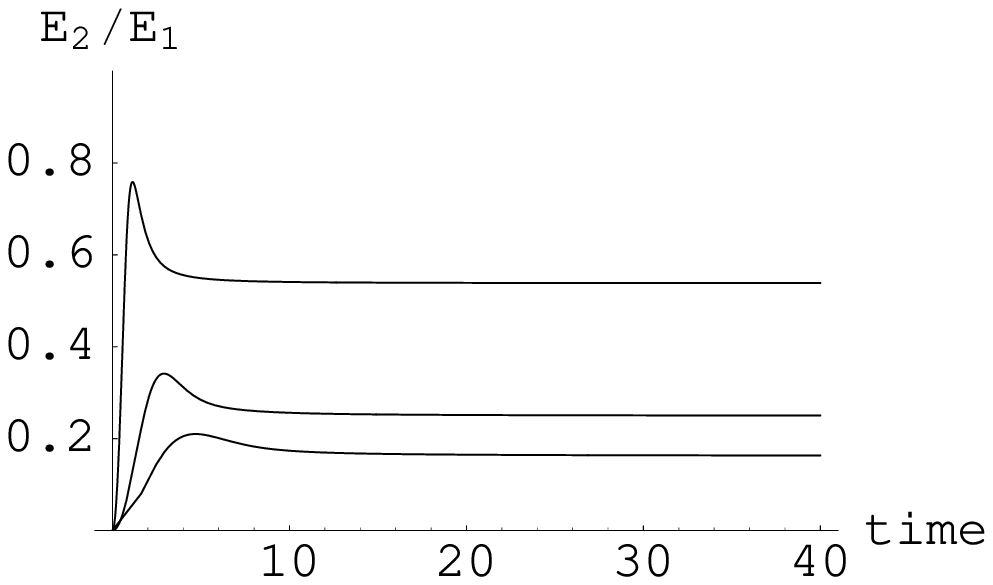}
\label{fig:2d_result}} \quad \subfigure[Data from N2, bins
  2,3,4 from top to
  bottom.]{\includegraphics[height=4cm]{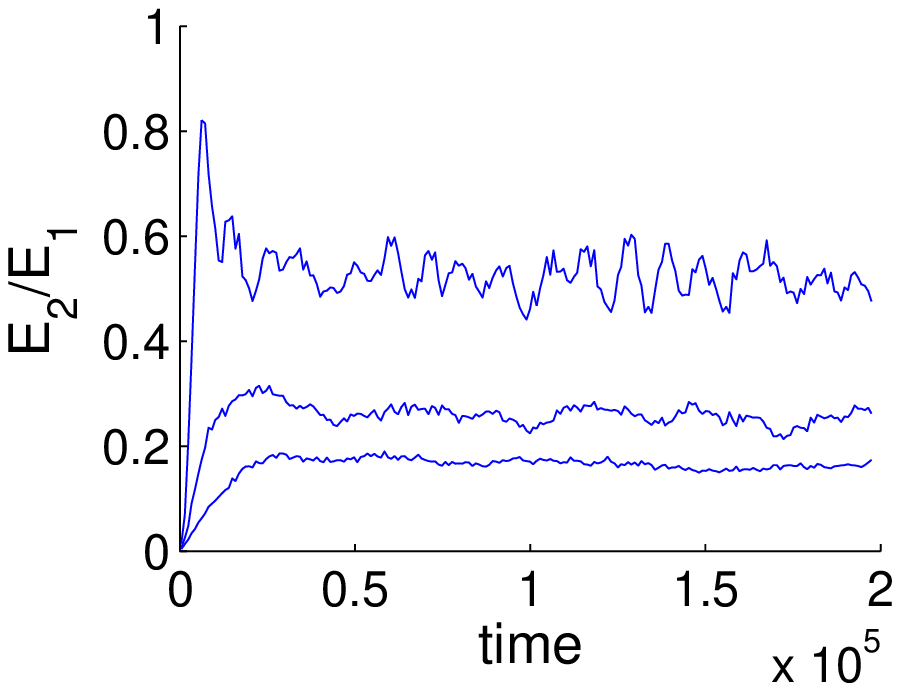}
\label{fig:num2_for_2d}}
  \caption{Comparison between the $2\times 2$ RMT model and the low
  frequency bins of N2 of Ref.~\cite{WL}. The energy ratio $E_2/E_1$ is plotted versus
  (dimensionless) time.} 
\label{fig:2d_and_num2}
\end{figure}

\subsection{Numerical simulations}
\label{sec:RMT_num}

The numerical simulations of the RMT model are performed as in
Refs.~\cite{GW,BGH}. The unfolding of the results, however, is not
done in the standard way. To compare with the numerical calculations
of Weaver and Lobkis, it is more convenient to first unfold the level
densities and then ``refold'' the spectra of our model with the
level densities of Ref.~\cite{WL}. Thereby we ensure that the 
mean-level densities of the RMT model acquire a form given by the Weyl
formula for the billiardlike system of Ref.~\cite{WL}. The
appropriate Weyl formula for the modal density in a square membrane
with Dirichlet boundary conditions reads~\cite{WL}
\begin{equation}
  \rho^{({\rm smooth})}(\Omega) = \frac{\Omega A}{2\pi c^2} -
                                  \frac{L}{4c} \ ,
\label{Weyl}
\end{equation} 
depending on the frequency $\Omega$. Here, $A$ is the area of the
membrane and $L$ is the side length of one room. Moreover, $c$ is
the wave velocity and is chosen as unity here. Using twice the area of a
room ($A = 2 \times 198^2$) and a value of $L$ which takes the roughness
of the boundaries into account ($L = 2 \times 3 \times 198$), we aquire an
expression for the level density of the entire system.  A comparison
with the Weyl formula of Ref.~\cite{berry_sinai} shows that the terms
from the extra edges introduced to model the disorder cancel. Hence,
we use the formula for a square membrane, which is precisely
Eq.~\eqref{Weyl}.  This Weyl formula is employed to refold our RMT
model for both numerical calculations N1 and N2. In the latter case,
the system is not really a billiard, due to the springs used for the
coupling. Nevertheless, the Weyl formula should be a good
approximation to the real level density.

For every simulation, we generate random matrices of dimension $2N
\times 2N=100 \times 100$, unfold, refold, calculate $E_2(t,\alpha)$,
and average over the results of 800 such simulations. As the functions
$E_2(t,\alpha)$ are not measured on the unfolded scale, but on that of
the Weyl formula, we do not introduce the notation
$\varepsilon_2(\tau,\lambda)$ in the present context.  Moreover, we
notice that the numerical calculations N1 and N2 depend on the mean
frequency $\Omega$ in the bin under consideration. Accordingly, we
arrive at a two-parameter family of time-dependent curves
$\overline{E_2}(t,\alpha,\Omega)$. We now associate each bin in the
numerical calculation performed by Weaver and Lobkis with the
corresponding frequency $\Omega$. This leaves us with a one-parameter
family of time-dependent curves $\overline{E_{2\mu}}(t,\alpha), \
\mu=1,\ldots,16$, for each bin labeled by $\mu$.

\subsection{Chaotically coupled regular systems}
\label{sec:RMT_pos}

In the RMT model defined by Eq.~\eqref{eq:RMT_model_def} and in its
subsequent analysis, we always drew the matrices $H_i, \ i=1,2$, from
(two independent) GOE's, implying that we model the two rooms
individually as chaotic systems --- before the coupling is considered.
One can also assume that the two rooms are regular systems before they
are coupled. In that case, the matrices $H_i, \ i=1,2$, would be drawn
from Poisson ensembles~\cite{GMGW}.  We also did such numerical
simulations. The results are qualitatively the same if the coupling --
i.e., the matrix $V$ -- introduces enough chaos. The main effect is an
adjustment of the scales. What matters for the qualitative behavior is
the chaoticity of the total system.

\section{Comparison with experiment and numerical calculations}
\label{sec:RMT_result}

\begin{figure}
\centering
\subfigure[Bin 2.]{
\includegraphics[height=4cm]{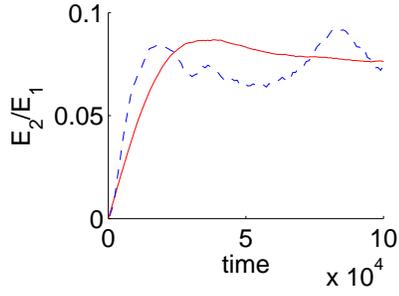}
\label{fig:RMT_N1_bin2}}
\quad
\subfigure[Bin 5.]{
\includegraphics[height=4cm]{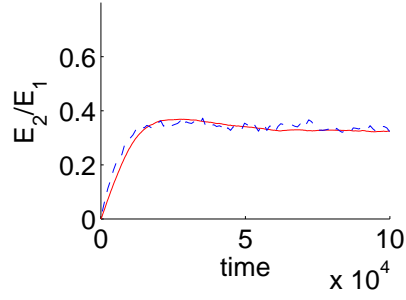}}
\\
\subfigure[Bin 9.]{
\includegraphics[height=4cm]{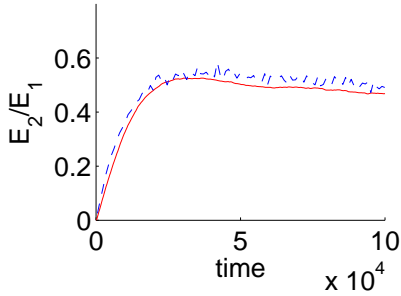}}
\quad
\subfigure[Bin 13.]{
\includegraphics[height=4cm]{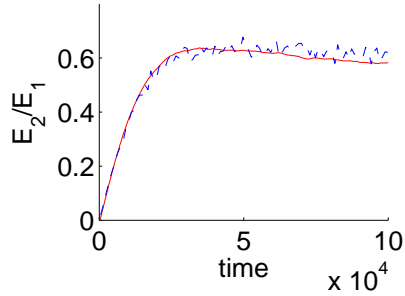}
}
\\
\subfigure[Bin 16.]{
\includegraphics[height=4cm]{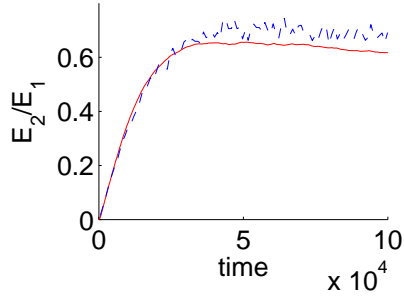}
\label{fig:RMT_N1_bin16}}
\quad
\caption{Energy ratio $E_2/E_1$ versus (dimensionless) time. Data from
  N1 of Ref.~\cite{WL} plotted as dashed lines and from refolded
  numerical simulations using the RMT model as solid lines. We notice the
  different $E_2/E_1$ scale in plot (a).}
\label{fig:RMT_result_N1}
\end{figure}

\begin{figure}
\centering
\subfigure[Bin 2.]{
\includegraphics[height=4cm]{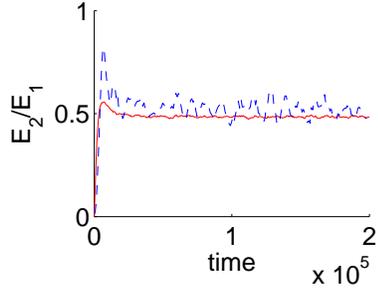}
\label{fig:RMT_N2_bin2}}
\quad
\subfigure[Bin 5.]{
\includegraphics[height=4cm]{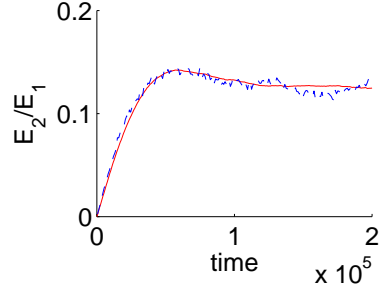}}
\\
\subfigure[Bin 9.]{
\includegraphics[height=4cm]{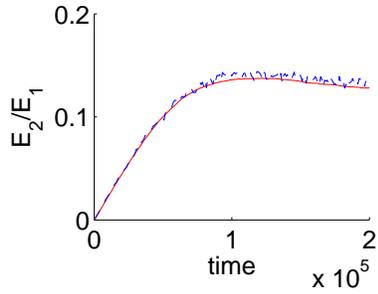}}
\quad
\subfigure[Bin 13.]{
\includegraphics[height=4cm]{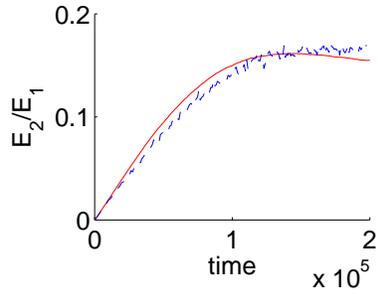}
}
\\
\subfigure[Bin 16.]{
\includegraphics[height=4cm]{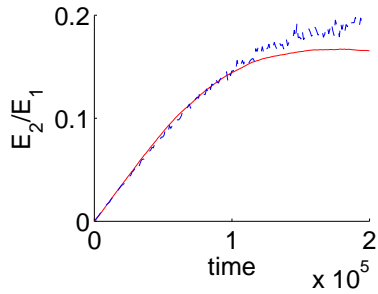}
\label{fig:RMT_N2_bin16}}
\quad
\caption{Energy ratio $E_2/E_1$ versus (dimensionless) time. Data from
  N2 of Ref.~\cite{WL} plotted as dashed lines and from refolded
  numerical simulations using the RMT model as solid lines. We notice the
  different $E_2/E_1$ scale in plot (a).}
\label{fig:RMT_result_N2}
\end{figure}

We now compare the results of our RMT model with those of
Ref.~\cite{WL}. The quantity studied in Ref.~\cite{WL} is mostly the
fraction of energy in room 2, or $E_2(t)/E$. In our figures, we plot
$E_2(t)/E_1(t)$ instead. We first consider the estimate,
Eq.~\eqref{eq:formfactor_result}, involving the GOE form factor and the
result~\eqref{e2est4} of the $2\times 2$ RMT model and compare with
the numerical calculations N1 and N2.  As formula~\eqref{e2est4} was
derived under the assumption of strong coupling on the scale of the
local mean-level spacing, it should apply to the high-frequency bins
of N1. Anticipating the later extraction of the coupling parameter, we
already now mention that indeed $\lambda>1$ in that bin. In
Fig.~\ref{fig:WL_num1_vs_form_factor} we compare Eq.~\eqref{e2est4}
with data from bin 16 (the highest frequency bin) of N1.  A good
description is obtained, although no equipartition is reached.  This
implies that the effective coupling $\lambda$ is large, but not very
large. We recall that formally $\lambda\to\infty$ corresponds to the
strongest coupling $\alpha=1$ -- i.e., to one single system.  Expansion
of the form factor for short times $\tau$ reveals a linear short time
dependence of the ensemble-averaged energy
$\varepsilon_2(\tau,\lambda)$. This is in agreement with the
analytical discussion of Ref.~\cite{WL}.

\begin{table}
\caption{Coupling parameters $\alpha$ for the $2 \times 2$ model used in
Fig.~\ref{fig:2d_result}. The asymptotic saturation $E_2/E$ values for
N2 in bins 2, 3, and 4 are taken from Ref.~\cite{WL}. The corresponding
coupling parameters are calculated from Eq.~\eqref{m7}.}
\begin{center}
\begin{tabular}{|c|c|c|}
\hline
bin & $E_2/E$ (asymptotic) & $\alpha$ \\
\hline
2 & 0.35 &  2.33 \\
3 & 0.2  &  0.67 \\
4 & 0.14 &  0.39 \\
\hline
\end{tabular}
\end{center}
\label{tab:2d_cplngs}
\end{table}

We turn to the $2\times 2$ RMT model and compare to bins 2, 3, and 4 of
N2. The values of the coupling parameter $\alpha$ are determined from
Eq.~\eqref{m7} and given in Table ~\ref{tab:2d_cplngs}, together with
the $E_2/E$ saturation values of N2. The $2\times 2$ RMT model curves
and data from N2 are shown in Figs.~\ref{fig:2d_result}
and~\ref{fig:num2_for_2d}, respectively. As expected, smaller values
of $\alpha$ correspond to a higher degree of localization -- i.e., to a
stronger deviation from equipartition.  The general behavior of the
results from the $2 \times 2$ model stays the same for large values of
$\alpha$. We notice that the time scale is different, because the
data from N2 were not unfolded.  The similarity shown in
Fig.~\ref{fig:2d_and_num2} between the $2\times 2$ RMT model and the
Weaver-Lobkis results~\cite{WL} is remarkable. This parallels the
success of $2\times 2$ RMT models for the spacing distributions.

The RMT simulations yield a one-parameter family of curves for each
bin, to be compared with the numerical calculations N1 and N2. We use
visual inspection to determine, for each bin, which curve fits best to
the numerical calculations N1 and N2. Typical results for some of the
bins are shown in Figs.~\ref{fig:RMT_result_N1}
and~\ref{fig:RMT_result_N2}. In the low-frequency bins of N2 there is
a discrepancy; see Fig.~\ref{fig:RMT_N2_bin2}. The RMT simulation does
not overshoot its saturation value as clearly as the data of
Ref.~\cite{WL}. This very large overshoot within a short-time interval
is, however, borne out in the $2\times 2$ RMT model; see
Fig.~\ref{fig:2d_and_num2}.  By the visual fit we determine the
coupling parameter $\alpha$.  We measure it on the scale of the local
mean-level spacing. In Table ~\ref{tab:num_RMT}, we list the in this
manner obtained coupling parameter $\lambda$ for N1 and N2 for the
bins under consideration. The $\lambda$ values extracted for N1 go up
with higher bins, because the saturation value comes closer to
equipartition. This effect is also visible for N2, but not so
pronounced, because the saturation value does not change much for
higher bins.

\begin{table}
\caption{Coupling constants $\lambda$ on the unfolded scale resulting 
from the RMT simulation, determined for some bins of the numerical
calculations N1 and N2 with mean frequencies $\Omega$: see
Figs.~(\ref{fig:RMT_result_N1}) and (\ref{fig:RMT_result_N2}).}
\begin{center}
\begin{tabular}{|c|c|c|c|c|}
\hline
bin & $\Omega$ & $\lambda$ for N1 & $\lambda$ for N2 \\
\hline
2 & 0.1562 & 0.4750  & 3.2316 \\
5 & 0.4686 & 2.0222  & 0.7754 \\
9 & 0.8851 & 3.0586  & 0.7502 \\
13 & 1.3016 & 3.9093  & 0.8441 \\
16 & 1.6140 & 4.2427  & 0.8899 \\
\hline
\end{tabular}
\end{center}
\label{tab:num_RMT}
\end{table}

We find good qualitative agreement between the RMT simulation and N1
and N2. Importantly, fluctuations between different random matrices
are quite large, and the RMT curves presented here are averages of
$E_2$ over 800 random matrices. Among those some show very much closer
similarity to the curves of Weaver and Lobkis~\cite{WL}, and we
believe that most of the discrepancy in the low-frequency bins of N1
are due to fluctuations resulting from the specific choice of the
boundary and perhaps other parameters in the numerical calculations of
Ref.~\cite{WL}. The reason for this interpretation of the discrepancy
is the peculiar form of the fluctuations in bin 2 of N1; see
Fig.~\ref{fig:RMT_N1_bin2}. Since these lower bins have a lower level
density, they should also be more sensitive to this kind of
fluctuations. In the intermediate-frequency bins, we find that the
numerical calculations N1 and N2 are well reproduced by the RMT model.
In the figures for the high-frequency bin, Figs.
\ref{fig:RMT_N1_bin16} and \ref{fig:RMT_N2_bin16}, the RMT results are
seen to deviate  
downwards for large times. It is conceivable that the short-time
behavior strongly depends on the specific realization of the coupling
while the long-time behavior is universal. 

The result from the experiment is shown in Fig.~\ref{fig:WL_exp}. The
qualitative features are well described by the RMT model.  It can be
seen that the system localizes in the low-frequency bins, but does not
do so in the high-frequency bins. In other words, the system
approaches equipartition of the energy. The localization behavior
depends on the strength of the coupling measured on the scale of the
local mean-level spacing, and the mean-level density is significantly
higher in the high-frequency bins. It seems that the saturation value
has not yet been reached on the time scale visible in
Fig.~\ref{fig:WL_exp}. A closer inspection shows an initial behavior
similar to the one in the results from our numerical RMT calculations
and a slight upwards trend of the energy ratio towards the end of the
time window studied. According to Ref.~\cite{WL} this means that,
first, the expected asymptotic value of $E_2/E_1$ is reached in the
middle of the time window studied and that, second, there is another
unknown effect acting on longer time scales adding to the energy spread
which is likely to be due to the coupling of the setup to the
environment.  Thus, we do not attempt to extract a quantitative
estimate of the coupling parameter $\alpha$ for the experiment.

\section{Conclusion}
\label{sec:dic_concl}

We have set up and analyzed an RMT model to describe the time behavior
of coupled reverberation rooms. This system shows localization effects
under certain conditions. Within our RMT model, we gave an estimate of
the strong coupling behavior which involved the two-level GOE
form factor. Moreover, we studied the $2\times 2$ version of our RMT
model analytically for arbitrary coupling strength and performed numerical simulations for the
$2N\times 2N$ version. 

{}From the comparison with the work of Weaver and Lobkis, we conclude
that the RMT model yields a good qualitative description.  Moreover,
we find an interpretation of the localization effect by relating it to
the universal features of RMT models for crossover transitions.

Formally similar RMT models have been studied in connection with
symmetry breaking. Then, the parameter $\alpha$ measures the degree of
symmetry breaking. This is so for isospin breaking in nuclear
physics~\cite{values,GW,26Al,BGH}, symmetry breaking in molecular
physics~\cite{Leitner}, and symmetry breaking in resonating quartz
crystals~\cite{cristais2}. The experiment that comes closest to the
present situation is the study of spectral correlations in coupled
microwave billiards~\cite{nos}. In this work and in the experiment of
Weaver and Lobkis~\cite{WL}, an excitation initially in system one
would stay there for all times if the coupling is zero. This is
formally analogous to a conserved quantum number.

In the RMT model the parameter measuring the size of the connection
has a most natural counterpart: the root-mean-square matrix element
due to the coupling measured on the scale of the local mean-level
spacing. Thus, there are two ways of making the effective,
dimensionless coupling parameter $\lambda$ small and to thereby
introduce localization: either the original coupling parameter
$\alpha$ which always has a dimension is small or the mean-level
spacing $D$ is made large.

It might be surprising that no equipartition of the energy is seen for
all coupling strengths even if one waits very long. This would be the
expectation if one compares the system with two water basins coupled
by a channel. Suppose that initially the two basins are empty and the
channel is closed by a gate. Now one of the basins is filled with
water and the gate is opened at $t=0$. Obviously, the water levels in
the two basins will be equal after sufficiently long time. The speed
with which this equipartition is reached simply depends on the cross
section of the channel. 

In the present case of the two coupled acoustic rooms the situation is
different, because the wave character of the excitations has to be
taken into account. Thus, the crucial parameter entering is the size
of the coupling connection -- i.e., its geometrical width -- compared to a
typical wavelength. The first waves after the excitation which come
from system 1 into system 2 enter a silent territory and cause the
first excitations there. The next waves coming from system 1,
however, encounter these first excitations in system 2 and interfere
with them constructively or destructively. This process continues and,
of course, after being excited in system 2, waves also travel back
into system 1. For smaller times, this complicated dynamics
certainly depends strongly on the realization of the coupling.  This
is clearly so in the numerical calculation of
Ref.~\cite{WL}. Nevertheless, the loneg-time behavior, in particular
the saturation limit, shows universal characteristics, consistent with
general features of quantum chaotic systems; see Ref.~\cite{GMGW}. In
the energy domain, this is borne out in the fact that the correlations
on smaller energy scales are described by universal RMT features,
while system-specific properties show up on larger scales, leading to
deviations from the RMT prediction. To avoid confusion, we emphasize
that these universal RMT features include those in the presence of
symmetry breaking.  By system-specific properties we mean, most
importantly, the scales set by the shortest periodic orbits.

It is worthwhile to realize that the chaoticity of the individual
subsystems before the coupling is not crucial, if the coupling itself
introduces enough chaos. We tested numerically that the behavior of
the corresponding model for two regular subsystems coupled chaotically
shows the same qualitative behavior. The saturation value is reached
slightly faster, which simply means that time is rescaled.  Finally,
we mention that our model is not restricted to elastomechanics. It
would also apply to coupled quantum dots and other coupled systems.

\section*{Acknowledgment}

This project was prompted by a conversation between R.L.~Weaver and
one of the present authors (T.G.) during a workshop at the Centro
Internacional de Ciencias (CIC/UNAM), Cuernavaca, Mexico. We thank
R.L.~Weaver for providing us with the data for the numerical
calculations N1 and N2. We acknowledge financial support from Det
Svenska Vetenskapsr\aa det.

\end{document}